\documentstyle[12pt]{article}
\def\singlespace {\smallskipamount=3.75pt plus1pt minus1pt
                  \medskipamount=7.5pt plus2pt minus2pt
                  \bigskipamount=15pt plus4pt minus4pt
                  \normalbaselineskip=15pt plus0pt minus0pt
                  \normallineskip=1pt
                  \normallineskiplimit=0pt
                  \jot=3.75pt
                  {\def\smallskip {\vskip\smallskipamount}}
                  {\def\medskip   {\vskip\medskipamount}}
                  {\def\bigskip   {\vskip\bigskipamount}}
                  {\setbox\strutbox=\hbox{\vrule
                    height10.5pt depth4.5pt width 0pt}}
                  \parskip 7.5pt
                  \normalbaselines}
\def\middlespace {\smallskipamount=5.825pt plus1.5pt minus1.5pt
                  \medskipamount=11.25pt plus3pt minus3pt
                  \bigskipamount=22.5pt plus6pt minus6pt
                  \normalbaselineskip=22.5pt plus0pt minus0pt
                  \normallineskip=1pt
                  \normallineskiplimit=0pt
                  \jot=5.825pt
                  {\def\smallskip {\vskip\smallskipamount}}
                  {\def\medskip   {\vskip\medskipamount}}
                  {\def\bigskip   {\vskip\bigskipamount}}
                  {\setbox\strutbox=\hbox{\vrule
                    height15.75pt depth6.75pt width 0pt}}
                  \parskip 7.25pt
                  \normalbaselines}
\def\dblspc {\smallskipamount=7.5pt plus2pt minus2pt
                  \medskipamount=15pt plus4pt minus4pt
                  \bigskipamount=30pt plus8pt minus8pt
                  \normalbaselineskip=30pt plus0pt minus0pt
                  \normallineskip=2pt
                  \normallineskiplimit=0pt
                  \jot=7.5pt
                  {\def\smallskip {\vskip\smallskipamount}}
                  {\def\medskip   {\vskip\medskipamount}}
                  {\def\bigskip   {\vskip\bigskipamount}}
                  {\setbox\strutbox=\hbox{\vrule
                    height21.0pt depth9.0pt width 0pt}}
                  \parskip 15.0pt
                  \normalbaselines}

\def\be{\begin{equation}}

\def\j-{\J_-}

\def\ee{\end{equation}}

\def\bearr{\begin{eqnarray}}
\def\bearrs{\begin{eqnarray*}}
\def\eearr{\end{eqnarray}}
\def\eearrs{\end{eqnarray*}}
\def\barr{\begin{array}}
\def\earr{\end{array}}

\def\nn8{\nonumber\\[10pt]}

\def\dis{\displaystyle}

\def\ed{\end{document}}

\begin{document}
\singlespace

\begin{center}
{\bf Beta decay rates for nuclei with 115 $<$ A $<$ 140 for r-process nucleosynthesis} 
\vskip 0.25cm

Kamales Kar$^{a}$, Soumya Chakravarti$^{b}$,
and V.R. Manfredi$^{c}$ \\ \ \\
$^{a}${\it Saha Institute of Nuclear Physics, 1/AF
Bidhan Nagar, Calcutta 700 064, India} \\
$^{b}${\it Department of Physics, California State Polytechnic University, Pomona, CA~91768, USA}\\
$^{c}${\it Dipartimento di Fisica ``G. Galilei", Universit\`a di Padova, Istituto Nazionale di Fisica Nucleare - Sezione di Padova, via Marzolo 8, I-35131, Padova, Italy} \\ 
\end{center}
\vskip 0.5cm

{\small
\noindent {\bf ABSTRACT:} 
For r-process nucleosynthesis the beta decay rates
for a number of neutron-rich intermediate heavy nuclei are calculated. The 
model for the beta strength function is able to reproduce the observed half~lives
quite well. 

\bigskip

Keywords: r-process, beta decay half life, Gamow Teller strength, waiting point nuclei

PACS classification no: 23.40.-s and 26.50.+x .

\bigskip


More than half of the nuclei heavier than iron are produced through r-process
nucleosynthesis \cite{burb-57}. Though one knows that high neutron density  
and fast neutron capture time scales compared to beta decay are required for
this process, astrophysical sites for them are not clearly identified. As the
r-process path often reaches highly neutron-rich nuclei, collecting experimental
information about them  sometimes becomes prohibitively difficult. On the other
hand near the neutron shell magic numbers depending on the shell gap \cite{mms-94}
and beta decay half lives the chain often has to wait long enough before
proceeding to the next heavier nucleus \cite{kratz-93}. These are called waiting point nuclei
and and can be identified by their observed relative abundances. With the lack of
experimental information, one often has to fall back on the predictions of 
theory for these neutron-rich nuclei. In this letter we consider such nuclei
near $N=82$ magic shell for $115<A<140$. 
The shell closure at $N=82$ for n-rich nuclei in the r-process path is
receiving a lot of attention lately and realistic models for beta decay
half-lives and rates in this region are badly needed. We construct here a model for the 
beta decay strength function based on the experience with lighter nuclei in
the fp shell. We deal with nuclei for which a few lowlying log ft's are known, to test the
model and then also apply  it to cases where not much is known experimentally
beyond the Q-value and half~life. The actual beta decay rates are also calculated for stellar
density and temperatures which are thought to be typical for the r-process.

Calculation of beta decay and electron capture rates have been pursued 
vigorously during the last twenty years for pre-supernova evolution and
gravitational collapse stage for core collapse supernovae \cite{lmp-03}.
Whereas for the sd-shell nuclei the rates are given by Oda {\it et al.} \cite{oda-94},
for the fp-shell nuclei for quite sometime the works of Fuller, Fowler and
Newman \cite{ffn-80} were used as the standard refrence. Aufderheide {\it et al.}
\cite{auf-91/94} indicated the usefulness of interacting shell model for the
rates. They also calculated approximate abundances in stellar conditions 
appropriate for the pre-supernova and collapse stage  and identified the nuclei which are the most important in
the network for each density-temperature grid point. On the other hand
statistical methods for the beta decay strength  distribution were used to
calculate the rates for fp-shell nuclei \cite{kar-94}. This was followed by large
space and detailed shell model calculations of the allowed beta decay strength
to calculate the weak interaction rates for nuclei with $A<65$ \cite{lmp-03}
\cite{lmp-01}. The effect of the new shell model rates on the late stage of
evolution of massive stars has also been seen \cite{heger-01}. Recently 
Pruet and Fuller \cite{pruet-03} used FFN-type ideas supplemented by 
experimental information wherever available to calculate the rates for nuclei
relevant for supernova collapse. However for much lower densities and 
temperatures appropriate for s- and r-process nucleosynthesis the work of 
Takahashi and Yokoi \cite{taka-87} is still being widely used. 
As large number of nuclei are involved in the r-process path the estimates of
half lives use a combination of global mass models and the quasi-particle 
random-phase approximation (QRPA). These are FRDM/QRPA \cite{moll-97} and
ETSI/QRPA \cite{bor-00}. Self-consistent Hartree-Fock-Bogoliubov plus
QRPA model have also been used \cite{engel-99}.

The beta decay rate for a nucleus in stellar conditions at temperaure T is
given by

\be
\barr{rcl}
\lambda= \dis\sum_{i}\; e^{-E_i/kT} \dis\sum_{j} \lambda_{ij} / Z
\earr
\ee

where $E_i$ is the energy of the state of the mother nucleus and Z is the
partition function of the nucleus. `$j$' sums over states of the daughter 
nucleus to which transitions are allowed. The rate from the parent nuclear
state `$i$' to the daughter nuclear state `$j$' is given by

\be
\barr{rcl}
\lambda_{ij}= \frac{\displaystyle \ln\ 2}{\displaystyle (ft)_{ij}} f_{ij}
\earr
\ee

 $f_{ij}$ is the phase-space factor for beta decay with an electron gas
outside \cite{kar-94}. For the Coulomb correction factor we use the Schenter-Vogel 
expressions \cite{schent-89} \cite{kar-94}. The allowed $ft$ values have two
components 

\be
\barr{rcl}
\frac{\textstyle 1}{\textstyle (ft)_{ij}} =\frac{\textstyle 1}{\textstyle (ft)_{ij}^{GT}} + \frac{\textstyle 1}{\textstyle (ft)_{ij}^{F}} = \frac{\textstyle 10^{3.59}}{\textstyle \mid M_{GT}\mid_{ij}^2} + \frac{\textstyle 10^{3.79}}{\textstyle \mid M_{F}\mid_{ij}^2}
\earr
\ee
 
For calculation of half~lives the decay is from the ground state, {\it i.e.}, there is
only one mother state with the Boltzmann factor equal to 1 in eq (1). 
In eq (3) the first part is the contribution from the allowed Gamow-Teller
operator and the second from the allowed Fermi operator. Normally
the GT strength distributes among three different types of final states of
the daughter :

(a) a discrete set of low-lying states for which the log ft's are known
experimentally

(b)  another set of discrete states above them for which the strengths are
not known 

(c)  a part of the GT giant resonance.

Usually one sees only the tail of the giant resonance as the Isobaric Analogue State (IAS) 
for  $\beta^-$ transitions is pushed up by the Coulomb interaction. In our treatment we
include in $\lambda$ of eq (1) as well as in the half~life the states of type (a), in particular
all the lowlying states connected by allowed transitions wherever known experimentally.
For the contribution from (c), {\it i.e.}, the giant resonance region, we adopt the procedure
detailed below.

The Fermi resonance is easy to take care of. In the absence of Coulomb
interaction, the total Fermi strength goes to IAS. Coulomb interaction spreads this
strength into a sharp resonance around the IAS and we take the resonance
width as $ \sigma_c = 0.157 Z A^{-1/3}$ \cite{kar-94} \cite{morita-73}. As
the spreading width is small this resonance cannot be reached by Q-value and
so Fermi makes very little contribution to half~lives and rates. But the 
situation for GT is quite different. As the spin-isospin dependent nuclear
Hamiltonian does not commute with the GT operator, the GT resonance is a
broad one. Detailed shell model results \cite{gomez-01} for fp shell nuclei and 
use of spectral distribution theory \cite{kotakar-89} 
indicate that the GT strength distribution to a good approximation can be taken
as a Gaussian for the initial state somewhat high in excitation (a few MeV)
to be in the chaotic regime. 
This point has been studied at length in the last few years. A good
understanding of the Gaussian forms follows by using Hamiltonians from
the two body random matrix ensembles which model chaos in many particle
quantum systems very well \cite{kota-01}.
Shell model results for lighter nuclei show that even with specific
two body interactions the Gaussian results persist. 
However for the ground state region there is some 
departure. In this letter we do not take into account such departure and take the
GT resonance for each nucleus given by a Gaussian. Hence the problem comes
down to fixing the centroid and width of the resonance. For the GT centroid
Bertsch and Esbensen \cite{bert-87} using the Tamm-Dancoff Approximation give the
following expression \cite{pruet-03}

\be
\barr{rcl}
     E_{GT^-} = E_{IAS}+ \Delta E_{s.o.} + 2 [ k_{\sigma \tau} S_{GT^-}/3 - (N-Z) k_{\tau} ]
\earr
\ee

where $\Delta E_{s.o.}$ is the contribution coming from the spin-orbit force.
For  $\Delta E_{s.o.}$ we use the orbit averaged value of 3.0 MeV \cite{bert-87}.
The last two terms come from the spin-isospin and isospin dependent nuclear 
forces. Here for the sum rule strength $ S_{GT^-}$ we use the expression $3 (N-Z) $. 
Following Pruet and Fuller \cite{pruet-03} we use the values $k_\tau = 28.5/A$ and $k_{\sigma\tau} = 23/A$.
For the overall normalisation of the sum rule Gaussian we use the quenched
 $3 (N-Z)$ value minus the experimental strength already included in (a). For
the quenching factor for the beta decay we use the value 0.6 \cite{kar-94}.
Finally to include the contribution from (b), {\it i.e.}, the discrete states below the
GT resonance region, we artificially broaden the resonance by using the width
 larger than the normal giant resonance widths. This is open to improvement.
 Perhaps partitioning of the space into subspaces and taking into account
a number of Gaussian strengths instead of a single Gaussian is a
better way of handling this. But then the width of each Gaussian either
has to be calculated theoretically or used as a parameter. 
We first use this model to calculate half~lives  of nuclei around
the $N=82$ shell closure. We include in this letter nuclei in the range
$115<A<140$ and also use the selection criterion that the Q-value greater
than 5 MeV for the statistical treatment of the high lying states. We separate
the nuclei into two groups - one set with known experimental ft values and the
other with no knowledge about low lying ft's.

We also mention here that the spectral distribution result for the
strength distribution gives it as a bivariate Gaussian in both the initial
and final energies. However for the r-process nuclei the stellar
temperature (in MeV) is quite low compared to the  energies
of the excited states of the mother. As a result the decay takes place
only from the ground state of the mother and hence the full bivariate
GT distribution does not come into play. 
In Table 1 we present the calculated half lives for 13 nuclei compared to their 
experimental values. These are obtained by taking the width of the GT giant
resonance as a free parameter and minimising the summed $ [ log (\tau^{cal}_{1/2}/\tau^{expt}_{1/2})]^2$.
The calculated values show reasonably good agreement with the observed values and 
$ \Sigma [ \log (\tau^{cal}_{1/2}/\tau^{expt}_{1/2})]^2 /N$ is 0.06 where `N'
stands for the number of nuclei. The optimised width is 5.0 MeV.

Table 2 shows a similar comparison for nuclei for which no lowlying log ft's
are known. Here the minimised $ \Sigma [ log (\tau^{cal}_{1/2}/\tau^{expt}_{1/2})]^2 /N$
is 0.07 for the width 7.7 MeV. The fact that the width for this case is larger
than the earlier one is expected as here the resonance has to be extended to 
the ground state region.

In Table 3 we calculate the  beta decay rates for $^{133}$Sn and $^{132}$In for the purpose 
of illustration over a range of density and temperature. The chemical potential
of the electrons changes from 1.32 MeV for the density $10^8$ g/cc to zero
for $10^3$ g/cc for the temperature $5 \times 10^8$ K. At higher temperatures
the chemical potential is less. For $3 \times 10^9$ K it is only 0.062 MeV for density 10$^8$ g/cc. We
have considered the density range far beyond the r-process to examine the 
dependence on density. In this range this dependence is found to be very mild but the
temperature dependence is seen to be a bit  stronger  for $^{133}$Sn. 
 Of course the rates can be calculated  for other nuclei and extended to other
densities and temperatures. As the values of $ kT$ are small compared to
the excitation energies of even the first excited state of the  mother 
nucleus we have given only the ground state contribution. For  similar
reasons the rates coming from `back resonances' of some special excited states with
large overlap with the daughter ground state region are also unimportant here and
hence are neglected.
In future we plan to extend these calculations to more neutron-rich nuclei
for which very little experimental information is available. We shall also consider 
heavier nuclei. We thank A.~Ray and M.M.~Sharma for encouragement.

\newpage

\begin{center} 

\begin{tabular} {rllrcll}
\hline \\
     &   Mother  &  Daughter  &  Q-value     & No of lowlying  &     $\tau_{1/2}$    &      $\tau_{1/2}$    \\
      &  nucleus   &  nucleus  &                      &  log ft's taken      &   Exp     &  Calc  \\
 & & & & & \\
 & & & (MeV) & & (sec) & (sec) \\ \\
\hline \\
 1.  & $^{138}_{53}$I$^{\ \ \ }_{85}$             &     $^{138}$Xe         &          7.820   &      3                    &      6.49 &  12.99          \\ \\

 2.  & $^{137}_{53}$I$^{\ \ \ }_{84}$         &     $^{137}$Xe            &           5.880  &     2                    &      24.5    &   33.1     \\   \\

 3.  & $^{136}_{53}$I$^{\ \ \ }_{83}$             &           $^{136}$Xe             &         6.930   &       3                  &      83.4   &  69.5 \\   

& & & & & \\

4.  & $^{134}_{51}$Sb$^{\ \ \ }_{83}$         &        $^{134}$Te       &          8.420  &      2                   &   10.43  & 7.72          \\  \\

5.  & $^{133}_{50}$Sn$^{\ \ \ }_{83}$         &        $^{133}$Sb       &          7.830  &      4                  &      1.44  &  1.19         \\  \\

6. & $^{132}_{51}$Sb$^{\ \ \ }_{81}$         &           $^{132}$Te        &        5.290    &       2                  &      167.4 &  141.3    \\    \\

7.  & $^{131}_{49}$In$^{\ \ \ }_{82}$           &       $^{131}$Sn         &         6.746  &      2                  &      0.282  &  0.280       \\  \\

8.  & $^{130}_{49}$In$^{\ \ \ }_{81}$           &           $^{130}$Sn     &       10.250  &      3                 &       0.32   &   0.25       \\   \\

9. & $^{128}_{49}$In$^{\ \ \ }_{79}$           &           $^{128}$Sn            &        8.980  &      3                 &       0.84   &  2.30       \\ \\

10. & $^{125}_{48}$Cd$^{\ \ \ }_{77}$         &           $^{125}$In             &         7.160    &      3                 &      0.65   &  2.12        \\ \\

11. & $^{120}_{47}$Ag$^{\ \ \ }_{73}$         &           $^{120}$Cd             &         8.200    &      4                 &      1.23   &  0.987        \\ \\

12. & $^{118}_{47}$Ag$^{\ \ \ }_{71}$         &           $^{118}$Cd            &         7.060  &      3                 &      3.76   &  7.23        \\ \\

13. & $^{116}_{45}$Rh$^{\ \ \ }_{71}$         &           $^{116}$Pd            &        8.900   &     3                  &      0.68   &  0.49        \\ \\

& & & & & \\
& & & & & \\

\end{tabular}

\end{center}

 

Table 1. Half-lives of nuclei for which $\log ft$ values are available and were used.

\newpage

\begin{center} 

\begin{tabular} {rlllll}
\hline \\
     &   Mother  &  Daughter  &  Q-value     &      $\tau_{1/2}$    &      $\tau_{1/2}$    \\
      &  nucleus   &  nucleus  &                      &   Exp     &  Calc  \\
 & & & (MeV) & (sec) & (sec) \\
\hline \\

1.  & $^{138}_{52}$Te$^{\ \ \ }_{85}$         &       $^{138}$I          &           6.370                    &      1.4  &  1.86         \\ \\

2.  & $^{137}_{52}$Te$^{\ \ \ }_{84}$         &       $^{137}$I          &           6.940                    &      2.49  &  1.41         \\ \\

3.  & $^{136}_{51}$Sb$^{\ \ \ }_{85}$         &       $^{136}$Te        &          9.300                     &      0.201  &  0.208          \\  \\

4.  & $^{132}_{49}$In$^{\ \ \ }_{83}$           &           $^{132}$Sn   &         13.6                     &      0.82 & 0.507        \\  \\

5.  & $^{117}_{45}$Rh$^{\ \ \ }_{72}$         &       $^{117}$Pd        &          7.000                     &      0.44  &  1.35          \\  \\

& & & & & \\
& & & & & \\
& & & & & \\
& & & & & \\

\end{tabular}
 
\medskip

Table 2. Half-lives of nuclei for which no $\log ft$ values are available. 

\end{center}

\newpage

\begin{center}

\begin{tabular}{llllll}

\hline \\
NUCLEUS &  DENSITY          &  \multicolumn{4}{c}{TEMPERATURE}    \\
                     &                             &                                                                        \\
                     & ( g/cm$^3$ )      &  \multicolumn{4}{c}{T$_9$ (in 10$^9$ K)}        \\
\\ \hline \\      
                     &                              &           0.5        &     1.0      &      2.0       &      3.0 \\ \\ \hline \\   
                     & 10$^8$               &           0.543    &   0.542   &    0.493    &      0.450 \\    
                     & 10$^7$               &           0.581    &   0.558   &    0.498    &      0.452 \\    
$^{133}$Sn& 10$^6$               &           0.584    &   0.559   &    0.498    &      0.452 \\    
                     & 10$^5$               &           0.584    &   0.560   &    0.498    &      0.452 \\   
                     & 10$^4$               &           0.584    &   0.560   &    0.498    &      0.452 \\    
                     & 10$^3$               &           0.584    &   0.560   &    0.498    &      0.452 \\ \\ \hline \\
                     & 10$^8$               &            1.70      &   1.75     &    1.59       &     1.46   \\
                     & 10$^7$               &            1.93      &   1.83     &    1.61       &     1.47   \\
$^{132}$In & 10$^6$               &            1.95      &   1.83     &    1.61       &     1.47   \\
                     & 10$^5$               &            1.95      &   1.84     &    1.61       &     1.47  \\
                     & 10$^4$               &            1.95      &   1.84     &    1.61       &     1.47  \\
                     & 10$^3$               &            1.95      &   1.84     &    1.61       &      1.47 \\ \\ \hline 

& & & & & \\
& & & & & \\
& & & & & \\
& & & & & \\
 
\end{tabular}

\medskip

Table 3. Decay rates in s$^{-1}$ for $^{133}$Sn and $^{132}$In. 

\end{center}

\newpage

\ed